\documentclass[a4paper,twocolumn]{esapub2005} 
\usepackage{subfigure}
\usepackage{times}
\usepackage{natbib}
\usepackage{graphicx}
\title{Monitoring of Transient Black Hole Candidates observed in the INTEGRAL survey}
\author[1,2]{F. Capitanio}
\author[1]{A. Bazzano}
\author[2]{A. J. Bird}
\author[1]{P. Ubertini}
\author[1]{M. Federici}
\affil[1]{IASF-roma INAF, Rome, Italy, Via Fosso del Cavaliere 100, I-00133 Rome, Italy}
\affil[2]{School of Physics and Astronomy, University of Southampton, Highfield Southampton, SO17 1BJ, UK}

\begin{document}
%
\maketitle
\begin{abstract}
 The INTEGRAL/IBIS survey was performed collecting all the GPS and GCDE data together with all the available public data . The second catalogue, published in 2006 by \citet{Bird1}, is dominated by detection of 113 X-ray binaries, with 38  being high--mass and 67 low--mass. In most systems the compact object is a neutron star, but the sample also contains 4 confirmed Black Holes and 6 LMXB black hole candidates (BHC). There are also, in additional, 6 tentative associations as BHCs based simply on spectral and timing properties. In the sample of 12 sources (BHC and tentatively associated BHC) there are 7 transient sources that went into outbursts during the INTEGRAL survey observations. We present here the monitoring of the time and spectral evolution of these 7 outbursts.
 \end{abstract}
 \vspace{-0.2cm}
 \section{BHC sample}
 \begin{itemize}
\vspace{-0.2cm}
\item {\bf IGR J17464-3213} is associated with H1743-322, a bright black hole candidate (BHC) observed by HEAO1 in 1977. After the start of the outburst in 2003, the source remained bright in soft X-rays (E $<$ 15 keV) for $\sim 8$ months 
and was regularly detected with the JEM-X monitor. At high energy, the IBIS telescope detected the source only on September 9 (52891 MJD). The source, after the first and most bright outburst, showed another two peculiar outbursts. The results of the data analysis of the first and brightest outburst, were published by \citet{Capit2}, \citet{Joinet}. A Radio-band observation campaign was also performed. A radio flare was seen by NRAO during the rising part of the first outburst (MJD=52728) and two jets were detected between the first and the second outburst (MJD= 52955--53049) \citep{Corb} For the second outburst, the analysed data show that the source had a shorter transition to hard state and then back to soft state with a peculiar temporal and spectral behaviour as the RXTE/ASM HR in figure \ref{IGR1746RXTE} shows \cite{CapB}. The data of the third outburst are still proprietary. Interestingly, the time period between the peak of an outburst and the subsequent one is equal and it has a value of about 400 days with a decreasing flux. This behaviour suggests a fourth outburst could occur in middle of September 2006.
\begin{small}
\begin{figure}[t!]
\centering
\includegraphics[width=0.6\linewidth,angle=90]{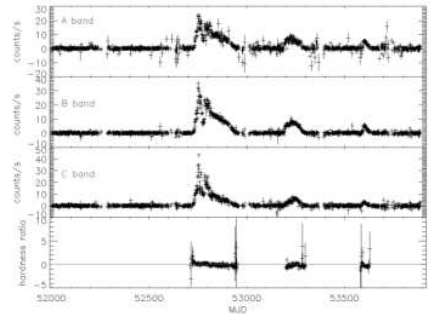}
\caption{IGR J17464-3212  RXTE/ASM light curves (energy bands: A-B-C) and the hardness ratio of the outburst periods (HR definition in the footnote n.1).\label{IGR1746RXTE}}
\end{figure}
\end{small}
\begin{small}
\begin{figure}[b!]
\centering
\includegraphics[width=0.6\linewidth,angle=90]{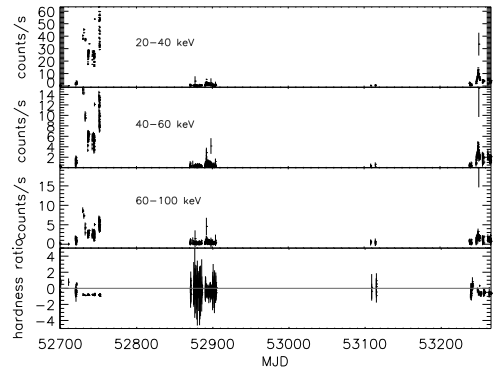}
\caption{IGR J17464-321 IBIS light curves and the corresponding HR (defined in the footnote n.2).\label{IGR1746RXTE_1}}
\end{figure}
\end{small}
Figure \ref{IGR1746RXTE} shows the RXTE/ASM light curves in three energy bands (A-B-C) and the hardness ratio (\footnote{$HR=({flux_{C}-flux_{A})/(flux_{A}+flux_{C}})$}{HR}). The three peaks of the source outbursts are clearly evident. The first peak was observed by INTEGRAL quite continuously (\citet{Capit1}), while the second peak was observed only in its declining part. Figure \ref{IGR1746RXTE_1} shows the IBIS light curves and the corresponding \footnote{$HR=({flux_{60-100}-flux_{20-40})/(flux_{60-100}+flux_{20-40}})$}{HR}, of all the INTEGRAL public observations of the source. 
The source spectral evolution of the first outburst followed the typical behaviour of a transient black hole: it was firstly detected in hard state, after passed thorough soft and very high soft state and, at the end of the outburst, it was back to hard state. Also the time evolution of the {\it flux vs photon index} diagram followed the "circular" or hysteresis-like behaviour expected for a transient BHC. Figure \ref{IGR1746eeuf_p} shows IBIS-JEMX spectrum of the Very high state of the source : 
the soft power law ($\Gamma = 3.1$) without any cut-off up to 200 keV is a signature of a non thermal comptonised emission spectrum \citep{Capit2}.
Figure \ref{IGR1746_1bispeak} (top panel) shows the JEMX-IBIS light curves in six energy bands of the declining part of the second outburst. It is noticeable the presence of a small peak, not evident in the RXTE one-day averaged light curves. The spectrum of this peak, that we show in the bottom panel figure, indicates a softening of the source with the presence of a moderate black body component and an hard tail without any cutoff up to 200 keV suggesting the presence of non thermal processes at work. 
\begin{small}
\begin{figure}[h!]
\centering
\includegraphics[width=0.9\linewidth]{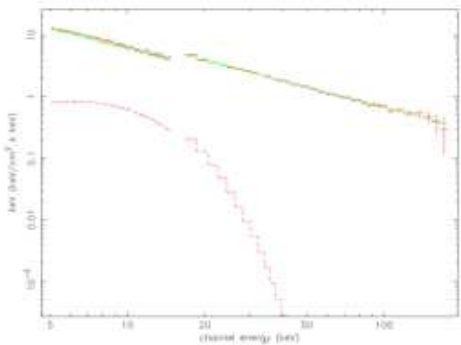}
\caption{ IGR J17464-3213 2003 outburst: Soft state at the very peak of the outburst. \label{IGR1746eeuf_p}}
\end{figure}
\end{small} 
\begin{center}
\begin{small}
\begin{figure}[t]
\centering
\includegraphics[width=0.8\linewidth]{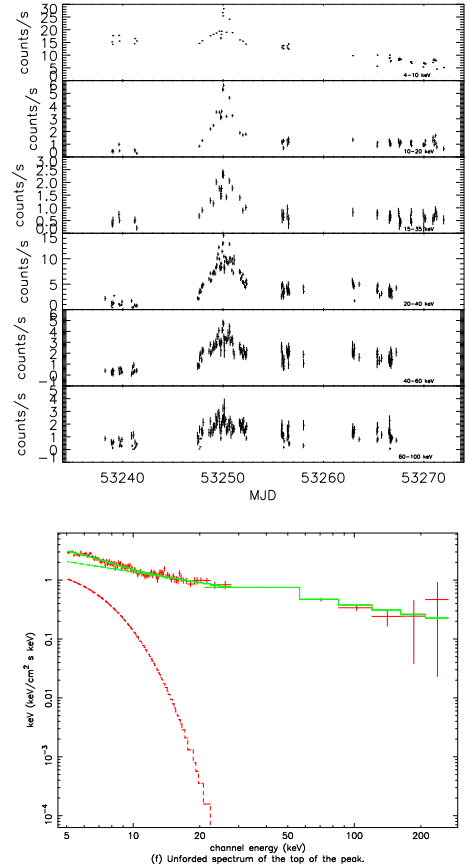}
\includegraphics[width=0.55\linewidth, angle=-90]{IGR17464spettro_top.ps}
\caption{IGR J17464-3213 second outburst. Top panel:JEMX (4-10, 10-20 and 20-35 keV) and IBIS (20-40, 40-80 and 80-100 keV) light curves of the final part of the outburst; bottom panel: IBIS-JEMX spectrum of a small secondary peak showed in the light curves (top panel).\label{IGR1746_1bispeak}}
\end{figure}
\end{small}
\end{center}
\vspace{-0.4cm}
\item {\bf IGR J17091--3624} was discovered  in 2003 April by {\it INTEGRAL}/IBIS
during its Galactic Centre Deep Exposure programme. Its flux reached 40 mCrab and 25 mCrab in the 15--40 keV and 40--100 keV  bands respectively. {\it RXTE\/} observed the source simultaneously on 2003 April 20, with an effective exposure of 2 ksec. Spectral and temporal evolution of the source, with a transition between the hard and soft states has been reported by \citet{Capit1}. Data collected by  {\it INTEGRAL\/} and {\it RXTE}  shows a Comptonised spectrum of the hard state as well as the JEM-X detection of a blackbody
component during the source softening. From archival data search, IGR J17091--3624 appears as a moderately bright transient source with flaring activity in 1994 October ({\it Mir}/KVANT/TTM), 1996 September ({\it BeppoSAX}/WFC), 2001 September ({\it BeppoSAX}/WFC \citep{Intz}), and 2003 April ({\it INTEGRAL}/IBIS, \citep{Kuu}).
IGR J17091--3624 is often below the detection
limits of IBIS and JEM-X in a single 2000-s SCW ($\sim$5 mCrab in the
20--100 keV IBIS energy range, and $\sim$23 mCrab in the 3--20 keV JEM-X energy
range). During MJD 52860--52924 (2003 August--October), the source was rather faint, especially in the 40--100
energy range, and the IBIS 20--40/40--100 keV hardness ratio shows a weak indication of softening corresponding to 25\% \citep{Capit1}. The 3 spectra in figure \ref{spe} represent the spectral evolution of the source during its outburst:\\
1) revs.\ 61--63 (2003 April 15--21, $\sim$15 ks), which is the first {\it INTEGRAL\/} observation combined with the {\it RXTE}/PCA data;\\
2) revs.\ 100--119 (2003 August--October, $\sim$77 ks), during which the IBIS spectrum softened, and which includes the only JEM-X detection.\\
3) revs.\ 165--179 (2004 April, $\sim$77 ks), containing only the IBIS data.\\
We clearly see a transition from the hard to the soft state from the first to the second epoch. Then, the source, after about one year, returns to the hard state, as shown by the IBIS data for revs.\ 164--179, (black spectrum). An intriguing feature of our hard-state spectra is the rather low electron temperature of the Comptonising plasma, $\sim$20 keV ($\tau \sim 2$). This low temperature is obtained in both {\sc compps} and {\sc comptt} models, and for both occurrences of the hard state, in the rise and decline. The detailed results on this source were published by \citet{Capit1}.
\begin{center}
\begin{small}
\begin{figure}[h!]
\centering
\includegraphics[width=0.65\linewidth,angle=90]{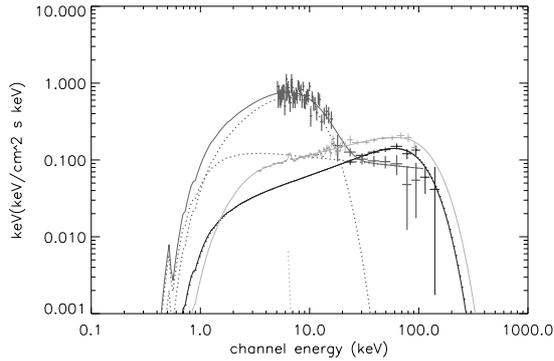}
\caption{Spectral evolution of IGR J17091-3624 during its outburst.\label{spe}}
\end{figure}
\end{small}
\end{center}
\vspace{-0.3cm}
\item {\bf XTE1908+094} has been detected for the first time during a RXTE/PCA scanning of the soft-gamma-ray repeater SGR 1900+14 \citep{Woods}
The source spectrum (2-30 keV) can be best fit with a power-law function including photoelectric absorption (column density $N_{h} $=$ 2.3\times 10^{22}$, photon index = 1.55).  An iron line
emission is present, but may be due to the Galactic ridge. The maximum source flux (2-10 keV) reached was from 64 mCrab, no coherent pulsations are seen between 0.001 and 1024 Hz. XTE J1908+094 is classified as a possible black hole candidate. Figure \ref{IGR1908} (top panel) reports the A,B and C RXTE/PCA bands light curves and the hardness ratio of the outburst periods. While the bottom panel shows the IBIS light curves in three energy bands; unfortunately the INTEGRAL data covers only the final part of the outburst, when the sources is coming back to hard state.
\begin{center}
\begin{small}
\begin{figure}[h!]
\centering
\includegraphics[width=0.6\linewidth,angle=90]{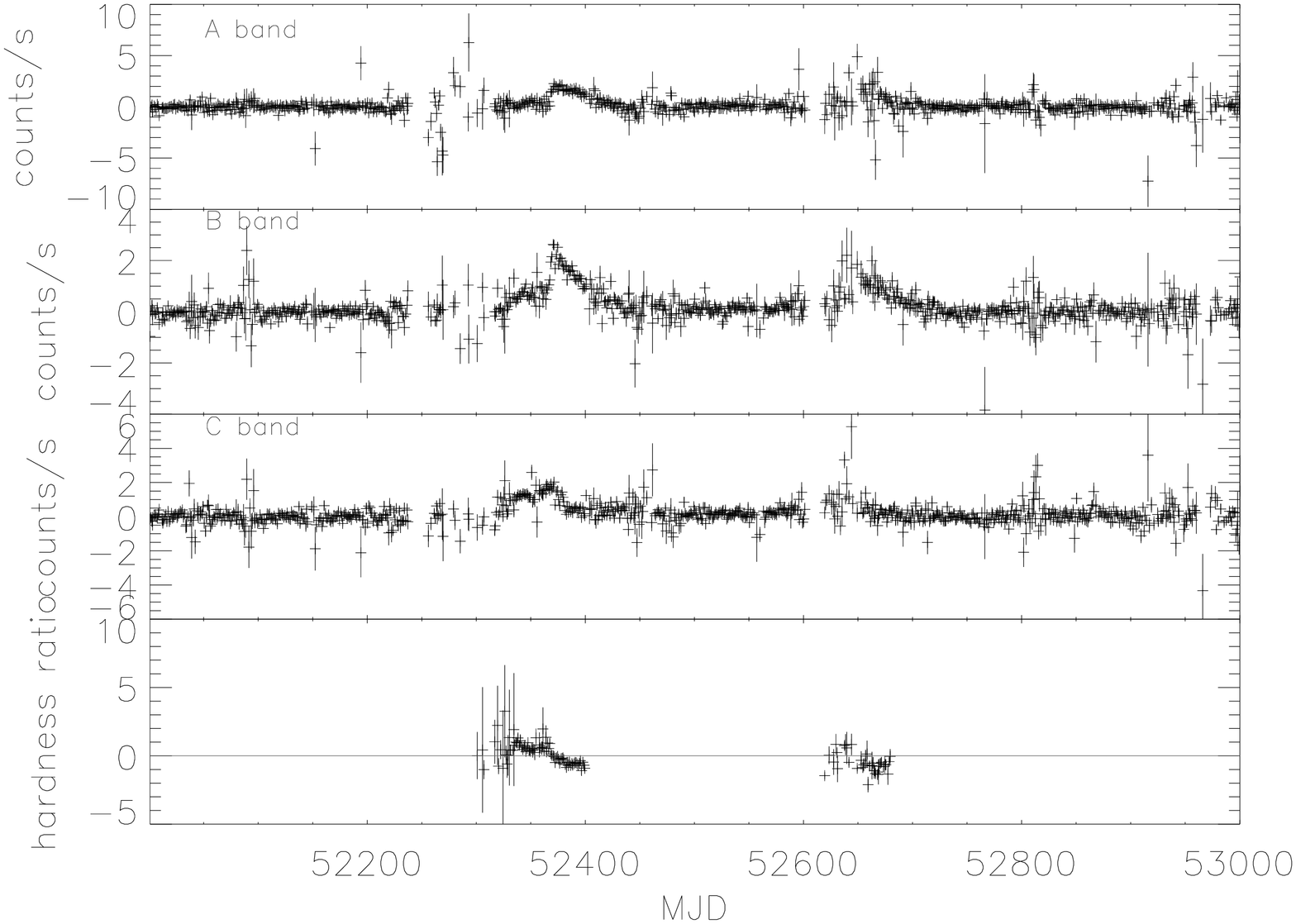}
\includegraphics[width=0.6\linewidth,angle=90]{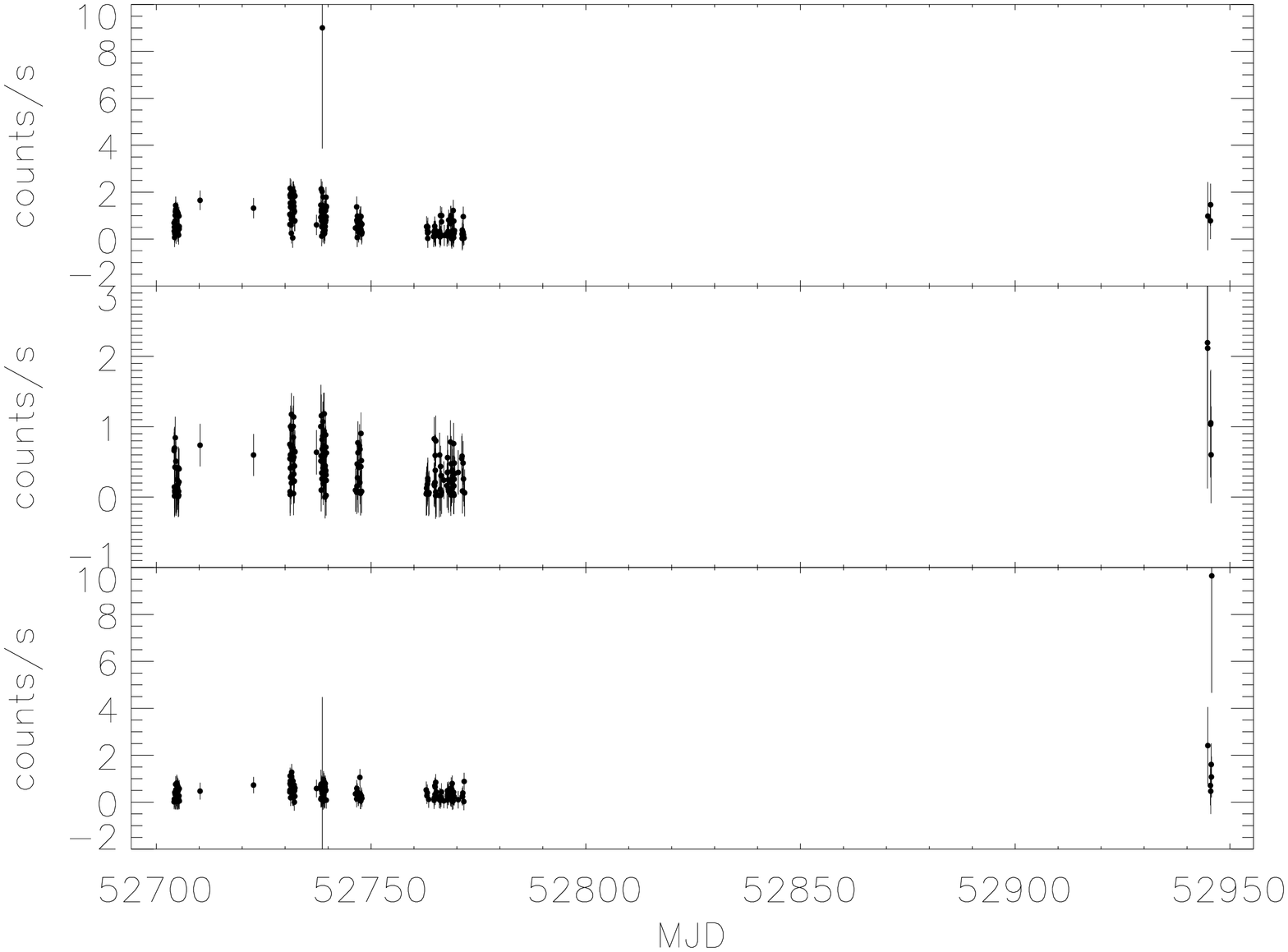}
\caption{XTE J1908+094. Top panel: A,B and C RXTE/PCA bands light curves and the hardness ratio; bottom panel: IBIS light curves in three energy bands (20--40, 40--60, 60-100 keV).\label{IGR1908}}
\end{figure}
\end{small}
\end{center}
%
%
 \vspace{-0.3cm}
 \item {\bf IGRJ18539+0727}, this faint transient source was detected during a routine scan of the Galactic Plane and deep observations of GRS1915+105 field on April 17-18 2003. It is classified as a probably BHC because of its hard spectrum. The spectral analysis of this source is still in progress. Figure \ref{IGR1839}  shows the  IBIS light curves and hardness ratio between the first and the second energy band. The HR shows a slight softening probably due to a spectral transition of the source.
 \begin{center} 
\begin{small}
\begin{figure}[h!]
\centering
\includegraphics[width=0.5\linewidth,angle=90]{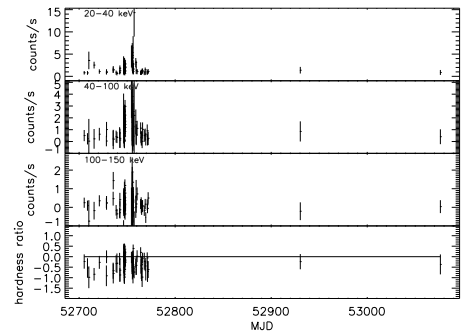}
\caption{ IGR J18539+0727: IBIS light curves in three energy bands (20--40, 40--60, 60-100 keV) and the hardness ratio.\label{IGR1839}}
\end{figure}
\end{small}
\end{center} 
\vspace{-0.3cm}
\item {\bf XTE J1720-318} was discovered on 2003 January 9 with the ASM monitor
onboard RXTE as a transient source in outburst. The source flux
increased to the maximum value of $\sim$430 mCrab in 2 days, and
then started to decay slowly. Follow up observations of
RXTE/PCA have shown the
presence of a 0.6~keV thermal component and a hard tail. The
spectral parameters and the source luminosity suggested a BHB in
 a High/Soft State. Soon after, a radio counterpart was identified with the VLA and ATCA radio
telescopes. 
XTE J1720-318 was observed by XMM-Newton, RXTE and INTEGRAL in
February during dedicated Target of Opportunity (ToO)
observations. It was then observed by INTEGRAL during the surveys
of the GC region performed in March and April and again from
August to October 2003. Even if the data coverage is not complete, the results obtained
from all these observations present a weak and steep tail in the hard energy range,
typical of a BHC in High/Soft State.
The source became active again at the end of March 2003. During this period
it showed a possible transition to hard state, and then decayed to a
quiescent state after April. In the hard state, the source was
detected up to 200~keV with a power law index of
$\sim$~1.9 and a peak luminosity of $\sim$~7~$\times $~$10^{36}$~erg~s$^{-1}$.
A detailed analysis of the RXTE1720-318 outburst was published by \citet{Cadolle}.
Figure \ref{XTE1720lc} (top panel) shows the RXTE-ASM light curves and the hardness ratio during the source activity periods, while the bottom panel of the same figure shows the IBIS light curves in three energy bands. 
The outburst of the source and its softening are clearly visible in both RXTE and IBIS light curves and hardness ratios. 
\begin{center}
\begin{small}
\begin{figure}[h]
\centering
\includegraphics[width=0.6\linewidth,angle=90]{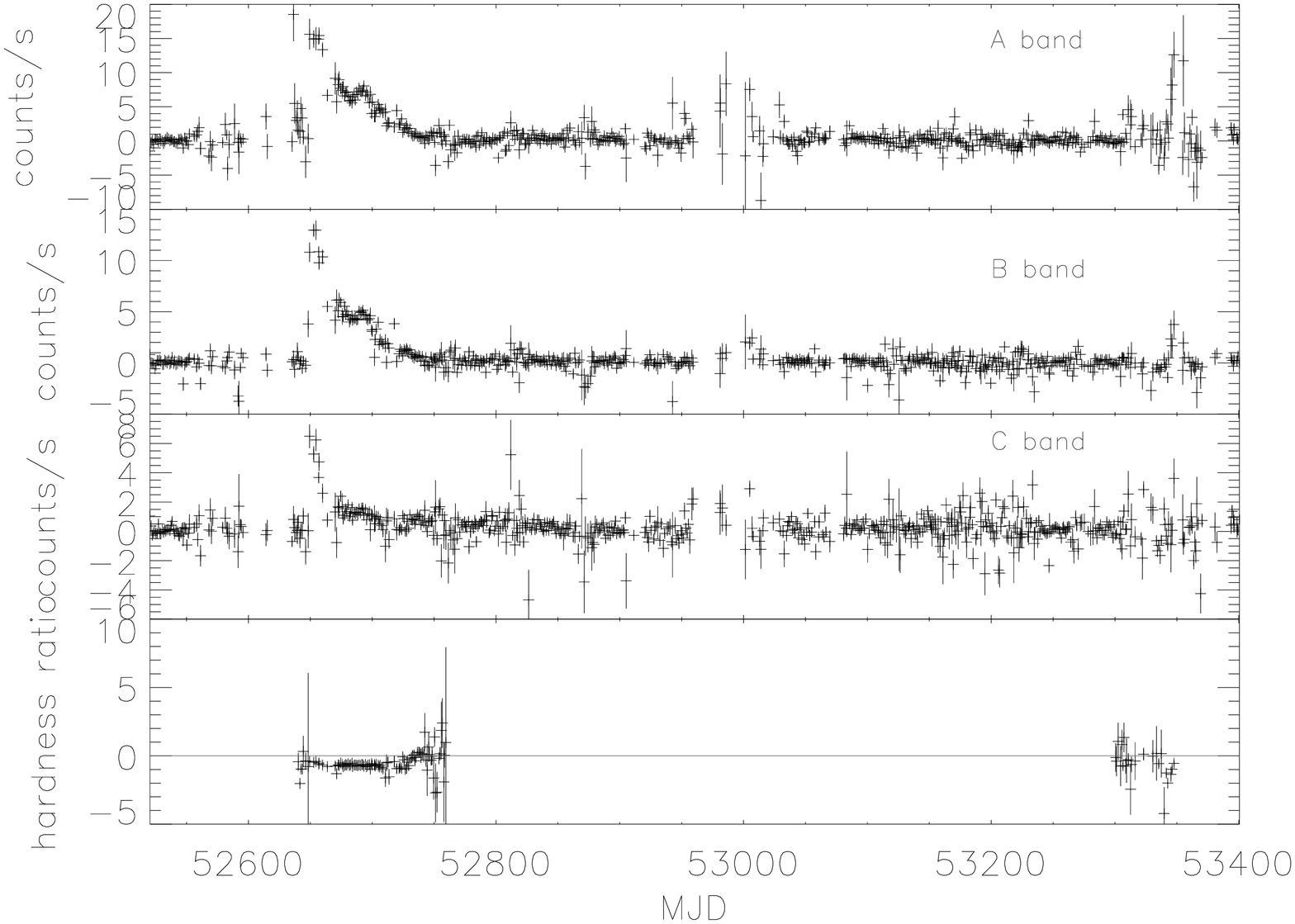}
\includegraphics[width=0.6\linewidth,angle=90]{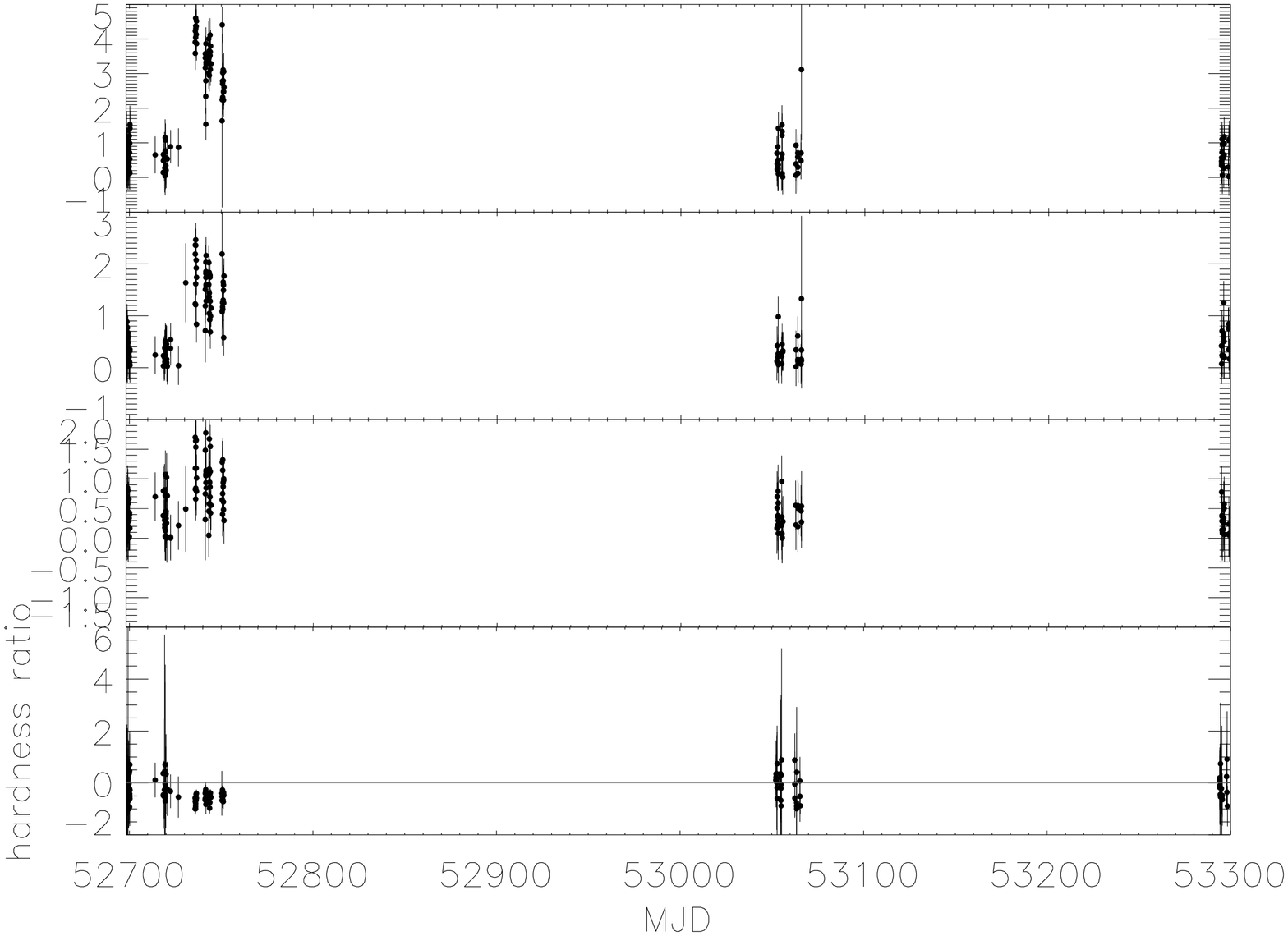}
\caption{XTE J1720-318. Top panel: A,B and C RXTE/PCA bands light curves and the hardness ratio; bottom panel: IBIS light curves in three energy bands (20--40, 40--60, 60-100 keV) and the hardness ratio\label{XTE1720lc}}
\end{figure}
\end{small}
\end{center}
 \vspace{-0.4cm}
 \item {\bf 4U1630-47} is one of the most active transient BHC. Its last outburst was very long covering more then two years (2002-2004).
  Moreover the source went again in outburst at the end of 2005. The 2002-2004 outburst, probably the longest observed for this source, was followed continuously by RXTE and INTEGRAL. 
 The analysis of INTEGRAL and RXTE 2002-2004 outburst data was published by \citet{Tom}.
 \vspace{-0.1cm}
 \item {\bf IGRJ17285-2922}  had a single period of activity in September 2003 when it was detected by INTEGRAL. The spectral characteristics and its position in the sky, very close to the galactic bulge, are consistent with a LMXB containing a black hole or a neutron star. Even if the nature of the compact object is still not constrained, its hard spectrum and the lack of type-I bursts suggests that the source could be a BHC. Figure \ref{IGR1785} shows IBIS light curves. A detailed analysis of the IGRJ17285-2922 data, collected by INTEGRAL, is published by \citet{Barlow}. 
 \begin{center} 
\begin{small}
\begin{figure}
\centering
\includegraphics[width=0.55\linewidth,angle=90]{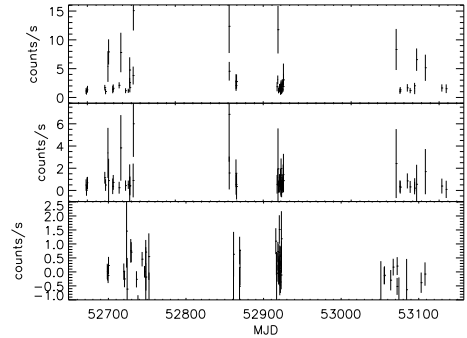}
\caption{IGRJ17285-2922: IBIS light curves in three energy bands (20--40, 40--60, 60-100 keV).\label{IGR1785}}
\end{figure}
\end{small}
\end{center} 
\end{itemize}
\vspace{-1.3cm}  
{\bf Acknowledgements}
 We acknowledge the ASI financial/programmatic support via contracts I/R/046/04.
We also thank Catia Spalletta for careful logistical support.
\begin{small}

\end{small}
\end{document}